\documentclass[12pt]{article}

\usepackage[margin=1in]{geometry}
\usepackage{ragged2e}
\usepackage{setspace}

\usepackage{titlesec}
\usepackage{graphicx}
\usepackage{amssymb}
\usepackage{siunitx}
\usepackage{multirow}
\usepackage{authblk}
\usepackage{url}

\usepackage[style=apa,backend=biber,sortcites=true]{biblatex}
\DeclareLanguageMapping{american}{american-apa}
\usepackage{caption}
\titleformat{\subsection}
  {\normalfont\itshape\normalsize}
  {\thesubsection}
  {1em}
  {}

\titleformat{\subsubsection}
  {\normalfont\itshape\small}
  {\thesubsubsection}
  {1em}
  {}

\title{Visual and Cognitive Demands of a Large Language Model-Powered In-vehicle Conversational Agent}

\author[1]{Chris Monk}
\author[1]{Allegra Ayala}
\author[1]{Christine S.P. Yu}
\affil[1]{Exponent, Inc.}

\author[2]{Gregory M. Fitch}
\author[2]{Dara Gruber}
\affil[2]{Google, Inc.}

\date{} % This hides the date

\begin{document}

%\onehalfspacing 
\maketitle

\begin{abstract}
\noindent
Driver distraction remains a leading contributor to motor vehicle crashes, necessitating rigorous evaluation of new in-vehicle technologies. This study assessed the visual and cognitive demands associated with an advanced Large Language Model (LLM) conversational agent (Gemini Live) during on-road driving, comparing it against hands-free phone calls, visual turn-by-turn guidance (low load baseline), and the Operation Span (OSPAN) task (high load anchor). Thirty-two licensed drivers completed five secondary tasks while visual and cognitive demands were measured using the Detection Response Task (DRT) for cognitive load, eye-tracking for visual attention, and subjective workload ratings. Results indicated that Gemini Live interactions (both single-turn and multi-turn) and hands-free phone calls shared similar levels of cognitive load, between that of visual turn-by-turn guidance and OSPAN. Exploratory analysis showed that cognitive load remained stable across extended multi-turn conversations. All tasks maintained mean glance durations well below the well-established 2-second safety threshold, confirming low visual demand. Furthermore, drivers consistently dedicated longer glances to the roadway between brief off-road glances toward the device during task completion, particularly during voice-based interactions, rendering longer total-eyes-off-road time findings less consequential. Subjective ratings mirrored objective data, with participants reporting low effort, demands, and perceived distraction for Gemini Live. These findings demonstrate that advanced LLM conversational agents, when implemented via voice interfaces, impose cognitive and visual demands comparable to established, low-risk hands-free benchmarks, supporting their safe deployment in the driving environment.
\end{abstract}

\RaggedRight
\setlength{\parindent}{20pt}

\section{Introduction}
Driver distraction remains a leading contributor to motor vehicle crashes and fatalities worldwide. The National Highway Traffic Safety Administration (NHTSA) reports that in the United States alone, distraction-related incidents claim thousands of lives annually (NHTSA, \citeyear{nhtsa2025}). The widespread use of mobile devices and increasingly complex in-vehicle technologies has amplified this challenge, as drivers are frequently tempted to engage with smartphones, infotainment systems, and other secondary tasks while driving \parencite[see][]{cox2023prevalence}. Extensive research, including meta-analyses and naturalistic driving studies, has consistently demonstrated that both visual-manual and cognitive distractions impair driving performance, but only visual-manual distractions elevate crash risk \parencite{caird2008meta,oviedo2016understanding,caird2018does,klauer2014distracted,fitch2013impact,victor2015analysis,dingus2019prevalence,fitch2022}.

The negative effects of distraction are multifaceted. Visual-manual distractions, such as looking away from the roadway or manipulating a device, are strongly associated with elevated crash and near-crash risk \parencite{klauer2006impact,fitch2013impact,victor2015analysis}. Beyond crash risk, visual-manual tasks have been shown to impair lane-keeping, increase reaction times to hazards, reduce speed control, and lead to more frequent and prolonged glances away from the road \parencite{horberry2006driver,harbluk2002impact,young2020using,mehler2016multi,chiang2005comparison}. These performance decrements can compromise vehicle control and hazard detection, especially in complex traffic environments. Cognitive tasks, which divert mental resources from driving, can compromise situational awareness and hazard responses \parencite{harbluk2002impact,horberry2006driver,regan2011driver,ranney2019detection,young2020using,strayer2013measuring,strayer2014measuring,strayer2015assessing,strayer2017visual,strayer2019visual}, although they do not necessarily increase crash risk and may offer drivers a method for counteracting underload or fatigue \parencite{dingus2019prevalence,gershon2009effects}.

Guidelines for minimizing driver distraction from built-in electronic displays, published by NHTSA and the Alliance of Automobile Manufacturers (now the Alliance of Automotive Innovation), were grounded in naturalistic driving evidence to establish clear limits on visual attention demand. Both organizations set an individual glance duration limit of two seconds as the upper threshold for acceptable glances away from the forward roadway (Alliance, \citeyear{alliance2006}; NHTSA, \citeyear{nhtsa2013}, \citeyear{nhtsa2014}). To further constrain the cumulative visual demand imposed by secondary tasks, the Alliance and NHTSA introduced a total eyes-off-road time (TEORT) criterion alongside the 2-second glance limit. The Alliance set the TEORT limit at 20 seconds (Alliance, \citeyear{alliance2006}), while NHTSA adopted a more stringent 12-second threshold (NHTSA, \citeyear{nhtsa2012}, \citeyear{nhtsa2013}). Although TEORT has not been directly linked to elevated crash risk \parencite{victor2015analysis}, it remains a key criterion in both sets of guidelines. In accordance with these recommendations, the present study utilized mean glance duration (MGD) and TEORT as primary measures of visual attention demand.

NHTSA’s guidelines were originally conceived as part of a three-phase approach to addressing distraction from in-vehicle systems (NHTSA, \citeyear{nhtsa2010}). The first phase focused on visual-manual distraction from built-in electronic devices (NHTSA, \citeyear{nhtsa2013}, \citeyear{nhtsa2014}), with subsequent phases intended to address portable and aftermarket devices (NHTSA, \citeyear{nhtsa2016}) and, lastly, voice-based interfaces. NHTSA has yet to publish any guidelines for voice-based interfaces. Though the initial guidelines were explicitly designed for visual-manual interactions, in the absence of dedicated criteria for voice-based interfaces, researchers have applied visual demand measures and thresholds from the visual-manual guidelines to voice-based tasks, as these remain the only established guidelines with an associated risk profile \parencite[e.g.,][]{monk2023visual}.

\subsection{Voice-based Interfaces}
Despite the establishment of guidelines to limit visual attention demand from built-in electronic displays, drivers continue to face distraction risks due to the proliferation of in-vehicle technologies (NHTSA, \citeyear{nhtsa2013}). While these guidelines have helped define acceptable boundaries for visual-manual interactions, NHTSA has yet to issue standards for voice-based systems, which are increasingly prevalent in modern vehicles. For example, intelligent personal assistants such as Amazon Alexa, Apple Siri, Google Assistant, and Microsoft Cortana are now integrated into many vehicles, enabling drivers to perform tasks, including navigation, communication, and entertainment, using voice commands rather than manual interactions \parencite{berdasco2019user}.

Empirical evidence supports the benefits of voice-based interfaces for reducing visual attention demand. Studies show that voice interactions lead to shorter mean glance durations and lower total eyes-off-road time compared to visual-manual interfaces \parencite{chiang2005comparison,mehler2016multi,shutko2009,monk2023visual}. For example, Monk et al. (2023) reported that a voice-based driving mode produced the lowest visual attention demand and subjective distraction ratings among several interface conditions, with glance metrics well below the recommended thresholds by industry (Alliance, \citeyear{alliance2006}) and NHTSA (NHTSA, \citeyear{nhtsa2013}, \citeyear{nhtsa2014}).

However, voice-based interfaces are not cognitively effortless. Certain tasks may still impose cognitive load, and voice commands can be error-prone in noisy environments \parencite{reimer2013preliminary,reimer2016multi}. While these systems reduce visual demand, they may increase cognitive load, which can impair hazard detection and driving performance \parencite{strayer2013measuring,strayer2014measuring,strayer2015assessing,strayer2017visual,strayer2019visual,dingus2019prevalence}.

\subsection{Effects of Conversations on Driver Performance and Behavior}

A substantial body of experimental and meta-analytic research demonstrates that engaging in conversation while driving, whether via handheld or hands-free phone, or with a passenger, produces measurable decrements in several aspects of driver performance. Meta-analyses and systematic reviews \parencite{caird2018does} consistently show that both phone and passenger conversations are associated with moderate increases in reaction time to hazards, slower detection of peripheral targets, and a reduction in the percentage of targets detected. These effects are robust across laboratory, simulator, and on-road studies.

Experimental work further clarifies the mechanisms underlying these effects. For example, both speech production (speaking) and comprehension (listening) during conversation draw on cognitive resources also needed for driving, leading to greater variability in vehicle control, especially under higher task demands or when the conversation requires active participation \parencite{kubose2006effects,rann2022effects}. Notably, interference is most pronounced during speech production, though comprehension also imposes a measurable load. These findings are echoed in recent cognitive modeling efforts that showed that conversation results in slower responses to secondary tasks and increased response caution \parencite{castro2023dynamic,tillman2017modeling}.

Importantly, the type of conversation, phone versus passenger, does not substantially alter the magnitude of these performance costs. \textcite{caird2018does} found that both forms of conversation produced similar effects on reaction time, detection, and lane-keeping. While some have hypothesized that passengers may help modulate conversation in response to traffic demands, the empirical evidence suggests that, on average, the cognitive demands of conversation are comparable regardless of whether the interlocutor is in the vehicle or remote. These findings are consistent with data showing that both hands-free phone and passenger conversations impair attention and peripheral detection, with no significant difference between the two types of interaction \parencite{amado2005effects}. More difficult conversations can further exacerbate these effects, particularly on peripheral detection tasks.

Despite these consistent laboratory and simulator findings, large-scale naturalistic driving studies provide a more nuanced and somewhat counterintuitive perspective on real-world safety outcomes. Analyses of naturalistic data on phone conversations while driving indicate that engaging in conversation, whether on a phone or with a passenger, was not associated with an increased risk of crash or near-crash events \parencite{dingus2019prevalence,fitch2013impact,hickman2012assessment,hickman2024cell,klauer2006impact,olson2009driver,victor2015analysis}. 

This apparent disconnect between experimental performance decrements and real-world crash risk is likely due to drivers’ adaptive behaviors. As noted by \textcite{strayer2016talking} and \textcite{castro2023dynamic}, drivers may compensate for the increased cognitive load of conversation by adjusting their driving, such as increasing following distance, reducing speed, or pausing conversation during complex maneuvers. These compensatory strategies may help mitigate the risks observed in controlled settings, resulting in no measurable increase in crash involvement in naturalistic studies.

\subsection{Measuring Distraction: Visual and Cognitive Dimensions}

Driver distraction encompasses visual, manual, and cognitive components \parencite{lee2008,regan2011driver}. Visual attention demand is typically assessed using eye-tracking or video coding of glance behavior to quantify glance locations and durations to use for testing tasks against glance criteria like those in NHTSA’s or the Alliance’s guidelines (NHTSA, \citeyear{nhtsa2013}, \citeyear{nhtsa2014}; Alliance, \citeyear{alliance2006}). Cognitive load, which is less directly observable, is often measured using the Detection Response Task (DRT; ISO 17488, \citeyear{ISO17488}), capturing changes in response time and miss rates during secondary task engagement \parencite[e.g.,][]{castro2016validating,strayer2019visual,monk2023visual,biondi2024adopting}. Subjective measures such as NASA-TLX and Likert scales complement these objective metrics \parencite{biondi2024using,monk2023visual}, though alignment between subjective and objective measures is not always perfect \parencite{kidd2010distracted,sanbonmatsu2013multi,von2022subjective}. \textcite{monk2023visual} found that voice-based driving modes reduced visual demand but did not consistently lower cognitive load compared to manual modes, highlighting the need for further research on cognitive load in voice-based and conversational systems.

\subsection{Conversational AI and Large Language Models in Vehicles}

The rapid evolution of large language models (LLMs) has fundamentally reshaped the landscape of voice interfaces, giving rise to conversational Artificial Intelligence (AI) systems that support multi-turn dialogue, contextual awareness, and adaptive, personalized responses \parencite{bond2025chatgpt}. Modern platforms such as Google’s Gemini Live and OpenAI’s ChatGPT exemplify this new generation of in-vehicle assistants, moving beyond the limitations of traditional voice systems that were restricted to single, command-based exchanges. These advanced agents can enable naturalistic conversation that allow for pauses, clarifying questions, and corrections which proactively engage users, offering a more natural and interactive experience \parencite{bond2025chatgpt}.

This transformation in human-machine interaction is underpinned by a growing body of theory and research in communication studies and human-AI relations. Historically, the “computers are social actors” (CASA) paradigm posited that people tend to treat computers and digital agents as if they were social beings, applying human social rules and expectations even when interacting with machines \parencite{reeves1996media,lee2024minding}. However, recent scholarship has begun to challenge and refine this view, noting that users’ responses to AI agents can differ from their reactions to humans, especially as AI systems become more sophisticated and human-like in both appearance and behavior \parencite{lee2024minding,henry-abion2025}.

Two key theoretical perspectives have emerged to explain these dynamics. The “mindlessness” account suggests that peoples’ social responses to machines are automatic and unreflective, triggered by minimal human-like cues. In contrast, the “machine heuristic” perspective argues that users rely on generalized beliefs or stereotypes about machines, such as expectations of objectivity, accuracy, or lack of emotion, which can shape their judgments and interactions with AI agents \parencite{lee2024minding}. Recent integrative models propose that users’ beliefs about machines, their prior experiences, and the context of interaction all influence whether they respond to AI as they would to another person, or whether they differentiate between human and machine sources \parencite{lee2024minding,henry-abion2025}.

The enhanced capabilities of conversational AI systems promise richer, more engaging user experiences in the vehicle. Unlike legacy voice assistants, these agents can maintain conversational context, adjust for input errors or delays while users think, adapt to user preferences, and facilitate complex, multi-step tasks. Theoretical frameworks such as the Mutual Theory of Mind \parencite{wang2022mutual} highlight the dynamic, reciprocal nature of human-AI communication, where both parties interpret and respond to each other’s cues in real time.

While the effects of conversation on driver performance have been well documented, the emergence of advanced conversational AI systems introduces new dimensions to the challenge of managing cognitive load and visual attention in the driving environment. Multi-turn dialogue with these agents requires drivers to maintain conversational context, interpret system prompts, and generate appropriate responses, all while actively monitoring the road. Theoretical frameworks in human-machine communication highlight that when AI agents take on more active, social roles, engaging in sustained, adaptive interactions, they may further elevate cognitive load and potentially draw attention away from driving \parencite{lee2024minding,henry-abion2025}.

Moreover, the anthropomorphic qualities of conversational AI, such as human-like speech, empathy, and adaptive behavior, not only enhance user engagement and foster trust and social presence but also introduce new challenges by blurring the boundaries between human and machine interaction. These features can potentially lead to over-reliance or increased distraction \parencite{henry-abion2025}. As a result, it is essential to examine how interactions with platforms like Gemini Live might impose cognitive and visual demands that differ from or exceed those of traditional voice interfaces or hands-free phone conversations. Understanding these dynamics is critical for evaluating the safety implications of deploying conversational AI like Gemini Live in vehicles and underscores the importance of selecting appropriate comparison tasks to accurately assess their impact on cognitive load and visual attention demand.

\subsection{Comparison Tasks}
To contextualize Gemini Live’s demands, the study incorporated three benchmark tasks that all maintained low visual attention requirements but spanned a wide range of cognitive load, from the relatively low demands of driving while following visual turn-by-turn guidance to the high cognitive demands of the operation span (OSPAN) working memory task. 

Visual turn-by-turn guidance was selected as a low-load baseline task to approximate “just driving” while ensuring participants followed route guidance during the experimental drive. Although the primary intent was to provide a minimal cognitive demand condition, navigation tasks inherently involve occasional visual and manual interaction, such as glancing at the display or confirming directions. Prior research indicates that simple navigation tasks typically impose modest workload compared to conversational or cognitively demanding secondary tasks, making them suitable as a practical baseline in on-road studies \parencite{mehler2016multi,reimer2013preliminary}. This approach allowed for a controlled comparison across task types while being generalizable to a real-world driving context.

Hands-free phone call was included as a comparison task because it represents a common form of voice-based interaction in vehicles. Importantly, large-scale naturalistic driving studies have consistently found that engaging in hands-free calls does not increase crash or near-crash risk. Analyses of real-world driving data indicate that drivers conversing on hands-free devices exhibit risk levels comparable to baseline driving conditions, in contrast to handheld phone use and visual-manual tasks, which significantly elevate crash risk \parencite{fitch2013impact,victor2015analysis,klauer2014distracted}. These findings suggest that while hands-free calls may impose cognitive demands, they do not translate into measurable safety degradation, making this task an appropriate benchmark for evaluating emerging voice-based technologies.

The OSPAN task was employed as the high-load anchor because it is a well-established and validated measure of working memory capacity and cognitive load. OSPAN requires participants to maintain and manipulate information while performing concurrent processing operations, creating substantial demands on executive resources \parencite{turner1989working,conway2005working,strayer2019visual}. This dual-task structure has been widely used in cognitive psychology and human factors research to induce high mental workload, making it an appropriate benchmark for the upper end of the cognitive demand spectrum in driving studies \parencite[e.g.,][]{unsworth2005automated,redick2012measuring,strayer2016talking,strayer2017visual}. Prior work has demonstrated that OSPAN performance correlates strongly with attentional control and task interference, reinforcing its suitability for assessing the impact of complex working memory challenges in multitasking environments \parencite{engle2002working,kane2003working,strayer2001driven}. 

\subsection{Present Study}
The present study was conducted on public roads to examine the effects of conversational AI and other common in-vehicle tasks on cognitive load, visual attention, and subjective ratings of workload and distraction. The study was designed to test the following hypotheses and address an exploratory research question.

The first hypothesis predicted that Gemini Live interactions and hands-free phone calls would impose cognitive load levels between those observed for visual turn-by-turn guidance (low load) and the OSPAN task (high load). Prior research has shown that hands-free phone calls increase cognitive demand relative to simple visual-manual tasks but remain well below the workload associated with complex working memory challenges such as OSPAN \parencite{strayer2013measuring,strayer2019visual,monk2023visual}. Similarly, Gemini Live interactions, particularly Multi-turn dialogue, were expected to result in cognitive loads similar to that from hands-free phone calls, which have established low risk associations \parencite[e.g.,][]{fitch2013impact,dingus2019prevalence}. 

The second hypothesis predicted that all five task conditions would result in MGDs below NHTSA’s 2-second threshold. Prior research consistently demonstrates that voice-based interactions substantially reduce visual distraction compared to manual tasks \parencite{mehler2016multi,monk2023visual}. Although minor differences in glance behavior were anticipated across conditions, all were expected to remain well within NHTSA’s criterion for individual glances.

Although TEORT is a widely used metric for visual attention demand inspired by guidance by NHTSA and industry efforts \parencite{nhtsa2013,nhtsa2014,alliance2006}, it is highly sensitive to overall task duration. For example, voice-based tasks that can continue for minutes will eventually exceed NHTSA’s 12-second criterion simply because of extended engagement rather than inherently unsafe glance behavior. However, voice-based tasks can include extended on-road glances that separate brief off-road glances, resulting in a glance pattern that differs from continuous visual-manual interaction. For this reason, no formal predictions were made regarding TEORT. However, given its transactional nature, the single-turn Gemini Live interaction was expected to produce the shortest TEORT values relative to other tasks because it involves a single query-response exchange rather than sustained interaction.

The third hypothesis predicted that subjective ratings of workload and distraction would correspond closely with objective measures of cognitive load and visual attention demand. Prior research has demonstrated general alignment between perceived workload and objective indicators of distraction \parencite{monk2023visual,von2022subjective}. Consistent with this pattern, participants were expected to report the highest perceived workload for the OSPAN task and the lowest for visual turn-by-turn guidance, with Gemini Live and hands-free phone calls rated similarly at acceptably low levels.

The exploratory question asked whether cognitive load changes over the course of extended interaction with Gemini Live Multi-turn, as a function of conversational length or elapsed time. Multi-turn dialogue introduces sustained engagement, raising the possibility of cumulative increases in cognitive demand as conversations progress. This analysis examined whether DRT miss rates or response times exhibited upward trends across successive conversational turns or with longer elapsed time.

\section{Method}\label{sec:method}

\subsection{Participants}
Thirty-two licensed drivers were recruited from the Bowie, MD area to achieve the target 24 participants. Six participants were excluded due to a mix of issues, including driving ability, early participation termination, and equipment failure. The study population was evenly balanced by sex and age, with 3 female and 3 male participants in age categories of 18-24, 25-39, 40-54, and 55-60. The participants reported driving at least one hour per week and never having been charged with driving under the influence (DUI). Participants were required to have owned a smartphone for at least 6 months and were required to have used a voice-based conversation agent, such as Google Assistant, Gemini Live, Siri, or ChatGPT Voice Mode, at least three times a week on their Android or home devices. They were paid \$225 for the study which lasted approximately two hours. A \$75 travel stipend was also offered to participants who lived more than a 30-minute drive from the Exponent’s Bowie, MD facility. This study was approved by Exponent’s Institutional Review Board.

\subsection{Test Environment and Measurement Equipment}
\begin{figure}[h!]
\begin{center}
\includegraphics[width=0.8\columnwidth]{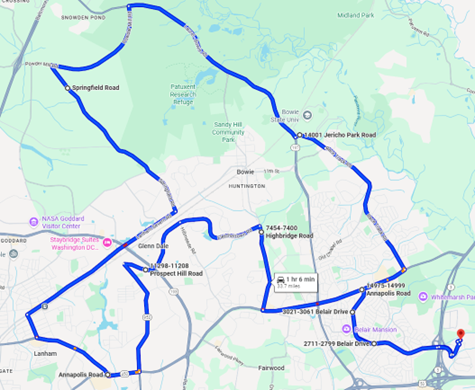}
\end{center}
\caption{Route in Bowie, MD.}
\label{fig:route}
\end{figure}

The test environment included driving a 2025 Chevrolet Trailblazer on a pre-determined route throughout Bowie, MD (Figure \ref{fig:route}). This route was planned through Google Maps, where pins were placed to minimize rerouting. The route took approximately one hour and 12 minutes to complete, and start times were selected to minimize traffic. The maximum speed limit along the route was 45 mph, highways were avoided, and stop signs and stoplights were kept to a minimum. If a participant deviated from the route, the experimenter instructed the participant to turn around and rejoin the designated route.

A Pixel 9 smartphone was utilized with both Bluetooth and wired connection for access to Android Auto and Google Maps for navigation. A pre-release version of Gemini Live in Android Auto was loaded onto the experimental phone. The phone was kept in the backseat for Experimenter access. All participants interacted with Gemini Live through the in-vehicle screen only.

\begin{figure}[h]
\begin{center}
\includegraphics[width=.8\columnwidth]{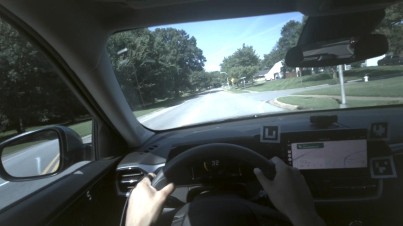}
\end{center}
\caption{View taken from the Tobii Pro 3's forward-facing camera.}
\label{fig:Tobiiex}
\end{figure}

Participant eye movement was recorded using Tobii Pro Glasses 3 head-mounted portable eye tracker, which recorded at 100 Hz \parencite{tobiimanual}. The forward-facing camera in the Tobii glasses was used to record participants’ point of view (see Figure \ref{fig:Tobiiex}). Data was processed through Tobii Pro Lab 25.7.

\begin{figure}[h!]
\begin{center}
\includegraphics[width=.8\columnwidth]{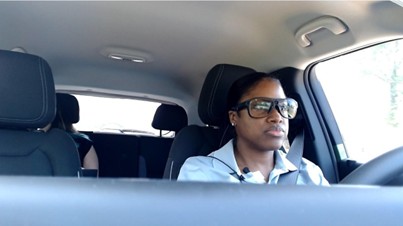}
\end{center}
\caption{An Exponent employee with the DRT tactor and Tobii Pro 3 Glasses.}
\label{fig:DRTex}
\end{figure}

Cognitive load was assessed using a tactile DRT system developed by Red Scientific, Inc. A vibrotactile transponder was positioned on the participant’s collarbone, over clothing if needed, and a microswitch was secured to their left middle finger using hook-and-loop fasteners (see Figure \ref{fig:DRTex}). The tactile stimulus lasted for 1 second, followed by a randomly varying interval between 2 and 4 seconds, resulting in a total cycle time of 3 to 5 seconds (ISO, \citeyear{ISO17488}). Participants were instructed to press the microswitch whenever they detected the vibrotactile signal. Data was recorded through Red Scientific’s RS Logger 1.3 software.

Subjective ratings of mental demand, physical demand, temporal demand, performance, effort, frustration, and three ratings of perceived distraction were taken with a 5-point Likert scale, with “1” representing “Strongly Disagree” and “5” being “Strongly Agree.” Participants’ ratings of satisfaction, ease of use, usefulness, clarity, and reliability when using Gemini Live were recorded for additional feedback. After completing the drive, participants were also asked to give examples of features they liked and disliked about Gemini Live as well as a rating of how likely they were to use Gemini Live on Android Auto again. Participants were also required to rank the experimental tasks in order of increased workload. Finally, participants were asked to rate their level of agreement using a 5-point Likert scale with statements, “Interacting with Gemini Live while driving required a lot of mental effort,” and “Interacting with Gemini Live while driving required the same mental effort as holding a hands-free cellphone conversation while driving.” They were then given the opportunity to provide open-ended feedback related to these two questions.

\subsection{Tasks}
Five experimental tasks were completed during testing, including the baseline driving condition with Visual Turn-by-turn Guidance only, Hands-Free Phone Call, Gemini Live Single-turn, Gemini Live Multi-turn, and OSPAN task. Visual turn-by-turn guidance was performed across all tasks using Google Maps, with voice navigation turned off to avoid interruptions to the Gemini Live and phone call tasks. The reliance on visual guidance alone may feel unnatural for drivers who regularly use voice-based navigation, potentially leading to increased glances to the screen as they adapt. Gemini Live Single-turn involved a single conversational interaction with the AI to complete a task, while Gemini Live Multi-turn required an extended dialogue with multiple back-and-forth exchanges, reflecting more complex conversational demands. 

The baseline task involving visual turn-by-turn guidance required participants to drive while following the pre-determined route displayed on the in-vehicle screen for a duration of five minutes. If a participant deviated from the route, they were instructed to navigate back to the route displayed on screen. The visual turn-by-turn guidance was present through all other task conditions. For the Hands-Free Phone Call task, participants were required to interact with the in-vehicle screen to call and converse with remotely located experimenters for approximately 1.5 minutes per trial. Participants were free to choose the topic for the call or reference one provided by the experimenter. The remote experimenter prompted participants to end the call when the allotted time had passed.

During task blocks where participants interacted with Gemini Live, the experimenter instructed them to alternate between Single-turn and Multi-turn tasks. Both tasks required that participants tap the in-vehicle screen to open the voice assistant and a second tap to enter Gemini Live mode. Participants were free to choose the topic for these interactions or reference experimenter-suggested prompts. During Single-turn tasks, participants were required to use prompts that elicited only one response from Gemini Live, after which interactions would be ended by tapping the on-screen button or a verbal “stop” command. Each Single-turn trial took approximately 30 seconds to complete. For Multi-turn prompts, participants were asked to continue to converse with Gemini Live, eliciting no less than two rounds of back-and-forth, until they felt satisfied with ending the conversation, which was expected to take an average of 3 minutes per trial. When ending either of these tasks, participants were given the option to manually end the conversation by pressing an on-screen button or to verbally end the conversation by saying “stop.” The total time spent on these tasks varied depending on several factors including the length of participant prompts and the length of Gemini Live’s responses.

Finally, the auditory OSPAN task required participants to complete 5 minutes of basic mathematical problems while recalling a series of single syllable words, which served as an anchor for high mental demand \parencite[see][]{strayer2016talking,strayer2019visual}. Specifically, each round of the OSPAN task included two steps: First, the in-vehicle experimenter provided verbal prompts such as “6/3 + 1 = 3” to which participants answered “true” or “false.” Second, after providing this answer, the participant would be given a monosyllabic word like “cob” or “bit.” After two to five rounds were presented, participants were asked to recall the words in the order that they had been given.

\subsection{Procedure}
Upon arriving at the experimenter facility, participants were introduced to the Chevrolet Trailblazer test vehicle and shown the route. Participants were instructed to adjust the driver seat and mirrors to their desired position before being instrumented with the eye tracking glasses and the DRT equipment. Participants were then instructed on all tasks while the vehicle was stopped, and before beginning any driving, participants were asked to complete practice rounds of the OSPAN task, Gemini Live tasks, and initiating the Phone Call task while stopped to ensure that they understood each task. Participants were then instructed on how to respond to the DRT stimuli and were reminded not to forget to respond for the entirety of the drive. Lastly, the eye tracker was calibrated by having participants stare at a target held at the distance of the steering wheel. Other instructions included maintaining the speed limit and following all traffic laws.

After the stationary practice rounds, participants were instructed to start driving along the route to initiate on-road practice rounds, with the first 5 minutes being turn-by-turn navigation to familiarize themselves with the vehicle. Participants then completed a practice round of each remaining task while driving. The duration of these practice rounds mirrored those of the testing rounds to prevent lack of familiarity from impacting results (see Table \ref{table1}). Participants then completed each task in the final testing while driving rounds. Task order was counterbalanced for each participant, with Gemini Live Single-turns and Multi-turns alternating within the same task block. Subjective feedback as described above was taken after each task. After participants concluded their tasks, they were instructed to continue following the route to end their drive by returning to the experimenter facility, where they answered the final round of subjective questions while parked. Finally, the eye tracker and DRT were removed, and participants were compensated before being excused.

\begin{table}[h!]
\caption{Study design with total estimated task times.}
\label{table1}
\begin{tabular}{llll}
\hline
Condition                                                               & Static Practice                                                             & Practice while Driving                                                        & Testing while Driving                                                         \\ \hline
\begin{tabular}[c]{@{}l@{}}Visual Turn-by-turn \\ Guidance\end{tabular} & None                                                                        & \begin{tabular}[c]{@{}l@{}}1 trial   \\ (5   min total)\end{tabular}          & \begin{tabular}[c]{@{}l@{}}1   trial   \\ (5   min total)\end{tabular}        \\ \hline
Hands-Free Phone Call                                                   & \begin{tabular}[c]{@{}l@{}}1   trial   \\ ($\sim$1 min total)\end{tabular}  & \begin{tabular}[c]{@{}l@{}}3   trials   \\ ($\sim$5   min total)\end{tabular} & \begin{tabular}[c]{@{}l@{}}3   trials   \\ ($\sim$5   min total)\end{tabular} \\ \hline
Gemini Live Single-turn                                                 & \begin{tabular}[c]{@{}l@{}}2 trials   \\ ($\sim$1   min total)\end{tabular} & \begin{tabular}[c]{@{}l@{}}6 trials   \\ ($\sim$3 min total)\end{tabular}     & \begin{tabular}[c]{@{}l@{}}3 trials   \\ ($\sim$3 min total)\end{tabular}     \\ \hline
Gemini Live Multi-turn                                                  & \begin{tabular}[c]{@{}l@{}}2 trials \\ ($\sim$6 min total)\end{tabular}     & \begin{tabular}[c]{@{}l@{}}6 trials\\ ($\sim$18 min total)\end{tabular}       & \begin{tabular}[c]{@{}l@{}}6 trials   \\ ($\sim$18 min total)\end{tabular}    \\ \hline
OSPAN                                                                   & \begin{tabular}[c]{@{}l@{}}2 trials   \\ (1 min total)\end{tabular}         & \begin{tabular}[c]{@{}l@{}}3   trials   \\ (5   min total)\end{tabular}       & \begin{tabular}[c]{@{}l@{}}3   trials   \\ (5   min total)\end{tabular}       \\ \hline
\end{tabular}
\end{table}

\subsection{Behavioral and Performance Data Reduction}
Data reduction and coding were conducted in R (version 4.5.2) via RStudio (version 2025.05.0) and Microsoft Excel (version 2502). See Table \ref{table2} for a summary of recorded and coded variables corresponding to each behavioral and performance measurements.

\subsubsection{DRT}

DRT variables recorded with Red Scientific’s RS Logger 1.3 software were exported, including task block and trial numbers, DRT trial numbers, DRT trial timestamps, and participant response to each DRT trial (i.e., whether participants responded to a given stimulus and, if so, their response time). Participant responses to DRT stimuli and response times were further used to code binary hit or miss outcomes. The outcome for a given DRT trial is considered a hit (1) if participants provided a response to trial stimulus and if participant responded no faster than 100 ms from stimulus onset but no later than 2500 ms from stimulus onset. All other types of response, including failure to respond to trial stimulus, or responding outside of the time 100-2500 ms window, were coded as misses (0).

\subsubsection{Eye Glance}
The forward-facing camera footage taken with Tobii glasses was processed in Tobii Pro Lab 25.7 to code intervals during which glances fell within the area of interest (i.e., Android Auto screen). The coding first entailed using Tobii Pro Lab gaze mapping to determine area of interest (AOI) hits, then, the assisted mapping feature was utilized to identify when participant gaze fell within the AOI using a similarity threshold of 80\% followed by manual review of each frame by the experimenters. Individual
glances lasting less than 100 ms were removed as they did not meet the criterion for minimum glance fixation duration specified in ISO 15700-1 (ISO, \citeyear{ISO15007}). The forward-facing camera footage was also used to identify intervals during which participants were bringing the vehicle to a stop (e.g., approaching and stopping at stop signs). Glances that occurred within these stopping intervals were excluded from analyses. Variables that were exported from Tobii Pro Lab include experimental task block and trial numbers, timestamps at 100 Hz frequency, and left and right eye identifiability. Glance data without at least one identifiable eye were excluded from analyses.

\subsubsection{Interactions with Gemini Live}
The forward-facing camera footage taken with Tobii glasses were exported from Tobii Pro Lab for video coding. The key variable of interest is the number of back-and-forth (i.e., conversational turns) that participants elect to have with Gemini Live during each trial of multi-turn conditions. During video coding, experimenters also noted down characteristics of interactions with Gemini Live unique to certain participants (e.g., anthropomorphizing Gemini Live) and situations that may preclude a trial from being included for analyses, including driving mistakes (e.g., missing a turn), technical issues (e.g., Gemini Live disconnection due to lack of service), or unusually challenging traffic situations (e.g., student driver ahead bringing their vehicle to a sudden stop mid-lane).

\begin{table}[h!]
\caption{Summary of driver behavioral and performance measurements, recordings exported, and variables coded.}
\label{table2}
\begin{tabular}{lll}
\hline
\begin{tabular}[c]{@{}l@{}}Behavior and \\ Performance\end{tabular} & Recorded Variables                                                                   & Coded Variables                                                                                                                                 \\ \hline
\multirow{5}{*}{DRT}                                                & Task block and trial number                                                          & \multirow{3}{*}{}                                                                                                                               \\ \cline{2-2}
                                                                    & DRT trial number                                                                     &                                                                                                                                                 \\ \cline{2-2}
                                                                    & DRT trial timestamp (ms)                                                             &                                                                                                                                                 \\ \cline{2-3} 
                                                                    & \begin{tabular}[c]{@{}l@{}}DRT trial responded (1) or \\ not (0)\end{tabular}        & \begin{tabular}[c]{@{}l@{}}Hit (1) = Responded (1) AND reaction \\ time between 100-2500 (ms)\end{tabular}                                      \\ \cline{2-2}
                                                                    & DRT trial reaction time (ms)                                                         & \begin{tabular}[c]{@{}l@{}}Miss (0) = Did not responded (0) OR \\ reaction time outside of 100-2500 (ms)\end{tabular}                           \\ \hline
\multirow{6}{*}{Eye Glance}                                         & \multirow{2}{*}{Tobii glasses camera footage}                                                & \begin{tabular}[c]{@{}l@{}}Whether vehicle is coming to a stop \\ (glances withing stopping interval \\ excluded from analyses)\end{tabular}    \\ \cline{3-3} 
                                                                    &                                                                                      & \begin{tabular}[c]{@{}l@{}}Whether glances are withing (1) or \\ outside (0) of AOI\end{tabular}                                                \\ \cline{2-3} 
                                                                    & Task block and trial   number                                                        &                                                                                                                                                 \\ \cline{2-2}
                                                                    & \begin{tabular}[c]{@{}l@{}}Timestamps at 100 Hz   \\ frequency (ms)\end{tabular}     &                                                                                                                                                 \\ \cline{2-3} 
                                                                    & \begin{tabular}[c]{@{}l@{}}Left eye identifiable (Valid \\ or Invalid)\end{tabular}  & \multirow{2}{*}{\begin{tabular}[c]{@{}l@{}}Glances within intervals where both \\ eyes are invalid were excluded from \\ analyses\end{tabular}} \\ \cline{2-2}
                                                                    & \begin{tabular}[c]{@{}l@{}}Right eye identifiable (Valid \\ or Invalid)\end{tabular} &                                                                                                                                                 \\ \hline
\begin{tabular}[c]{@{}l@{}}Gemini Live \\ Interactions\end{tabular} & Tobii glasses camera footage                                                          & \begin{tabular}[c]{@{}l@{}}Number of conversational turns \\ Participant characteristics \\ (e.g., anthropomorphizing)\end{tabular}             \\ \hline
\end{tabular}
\end{table}

\subsection{Data Analysis}

Any test trials that were precluded from analyses, as identified during data reduction, were substituted with practice trials for the same task completed while driving. The DRT and glance data were analyzed with repeated measures ANOVAs using linear mixed models. This approach is particularly well suited for unbalanced designs with different numbers of observations across participants. All models specified Task as a fixed effect, and participant as a random effect. Post-hoc pairwise t-test comparisons with the Holm adjustment and Satterthwaite’s method for degrees of freedom \parencite{kuznetsova2017b} were used, along with Cohen’s $d$ for effect size. For each model, the marginal $R^2$, which describes the proportion of variance explained by the fixed factor alone, and conditional $R^2$, which describes the proportion explained by the fixed and random factors, are reported. The objective measures plots show estimated marginal means and 95th percentile confidence intervals. 

All data analyses and visualizations were conducted in R (version 4.5.2). The linear mixed-effects models were fitted using the R package lme4 \parencite[version 1.1-37;][]{bates2015fitting}; parameters in the model were tested using the package lmerTest \parencite[version 3.1-3;][]{kuznetsova2017b}. Paired t-test comparisons with the holm adjustment and Satterthwaite’s method for degrees of freedom, as well as Cohen’s $d$ were calculated using the emmeans package \parencite[version 2.0.0;][]{lenth2025emmeans}.

\section{Results}\label{sec:result}
See Table \ref{table3} for pairwise comparison results for DRT Miss Rate, DRT Response Time (RT), Mean Glance Duration (MGD), and Total Eyes Off Road Time (TEORT).

\subsection{DRT Miss Rate}
\begin{figure}[htbp!]
\begin{center}
\includegraphics[width=1\columnwidth]{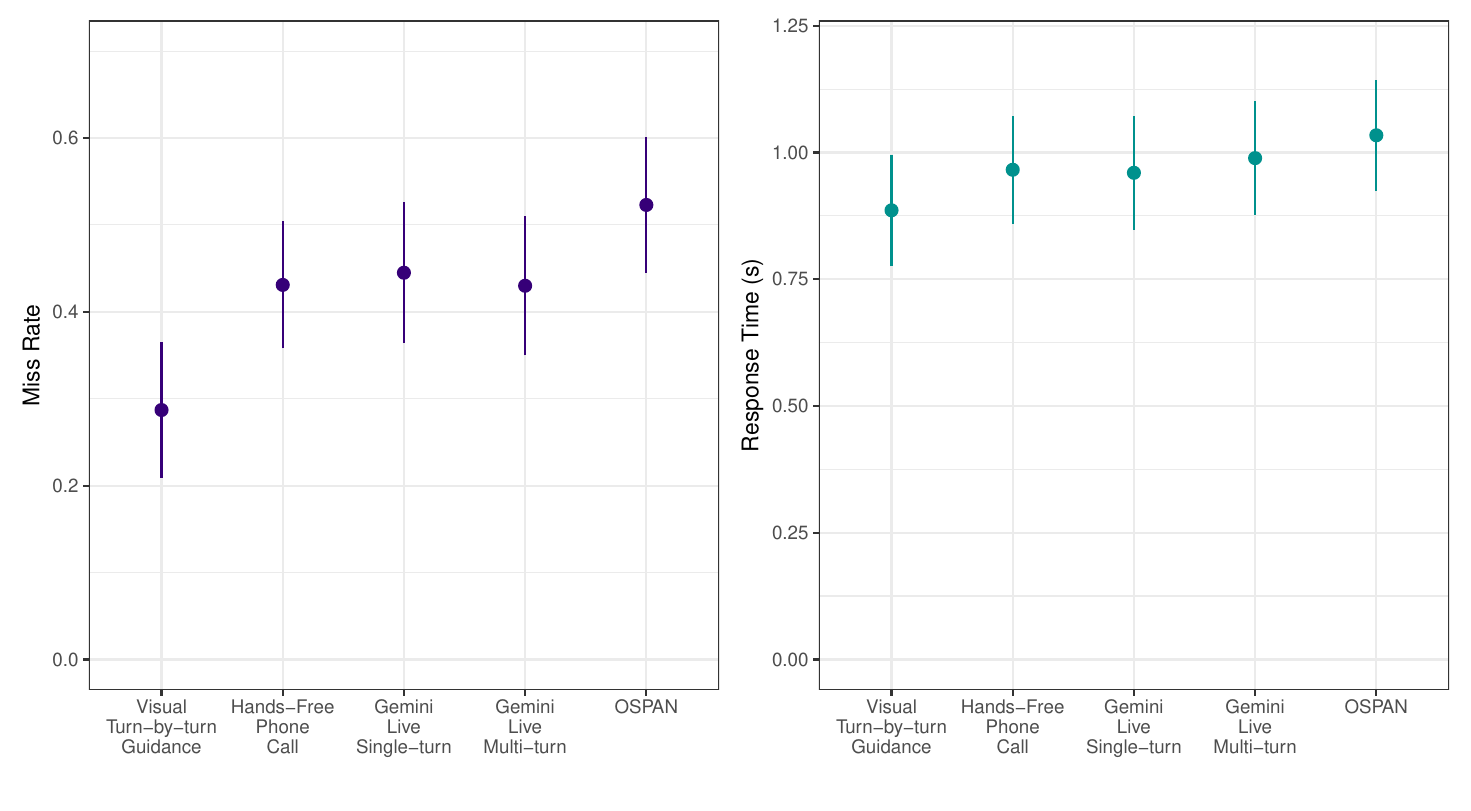}
\end{center}
\caption{DRT estimated marginal means and 95\% Confidence Intervals.}
\label{fig:DRTdata}
\end{figure}

Results of the repeated measures ANOVA on the DRT miss rate linear mixed model showed a significant effect for Task, $F$(4, 633.45) = 15.89, $p < $ .0001, partial $\eta^2$ = 0.09. The marginal $R^2$ for the miss rate model was 0.065 and conditional $R^2$ was 0.348. Post-hoc paired comparisons showed significant differences in Miss Rate across task conditions, reflecting variations in cognitive load as seen in Figure \ref{fig:DRTdata}. Visual Turn-by-turn Guidance yielded the lowest miss rate ($M = $ 0.287, $SE = $ 0.039), which was significantly lower than Hands-Free Phone Call ($M = $ 0.431, $SE = $ 0.036), $t$(634.51) = -5.42, $p < $ .0001, $d = $ 0.63, Gemini Live Single-turn ($M = $ 0.445, $SE = $ 0.040), $t$(632.38) = -4.86, $p < $ .0001, $d$ = 0.69, Gemini Live Multi-turn ($M = $ 0.430, $SE = $ 0.040), $t$(632.76) = -4.52, $p$ = .0001, $d$ = 0.63, and OSPAN ($M = $ 0.523, $SE$ = 0.038), $t$(632.71) = -7.83, $p < $ .0001, $d$ = 1.04. Furthermore, both Hands-Free Phone Call and Gemini Live Multi-turn conditions resulted in significantly lower miss rates than the OSPAN task (Hands-Free Phone Call vs. OSPAN: $t$(631.90) = -3.54, $p$ = .0026, $d$ = 0.40; Gemini Live Multi-turn vs. OSPAN: $t$(633.51) = -2.94, $p$ = .0171, $d$ = 0.41). These results indicate that while all voice-based and high cognitive load tasks increased cognitive demand relative to visual turn-by-turn guidance, the OSPAN task imposed the greatest cognitive load overall. Notably, both Gemini Live interaction modes and hands-free phone call produced similar levels of cognitive demand, higher than visual turn-by-turn guidance but lower than the OSPAN benchmark.

\begin{table}[h!]
\caption{Summary of pairwise comparison results for Miss Rate, DRT Response Time (RT), Mean Glance Duration (MGD), and Total Eyes Off Road Time (TEORT)}
\label{table3}
\begin{tabular}{lcccccccc}
\hline
                                                                                                 & \multicolumn{8}{c}{Measures}                                                                                 \\ \cline{2-9} 
                                                                                                 & \multicolumn{2}{c}{Miss Rate} & \multicolumn{2}{c}{RT} & \multicolumn{2}{c}{MGD} & \multicolumn{2}{c}{TEORT} \\ \cline{2-9} 
Comparison                                                                                       & t              & d            & t           & d        & t           & d         & t            & d          \\ \hline
\begin{tabular}[c]{@{}l@{}}Visual Turn-by-turn Guidance\\ - Hands-Free Phone Call\end{tabular}   & -5.42$^{\ddagger}$         & 0.63         & -2.65       & 0.32     & -0.76       & 0.18      & 2.38         & 0.56       \\
\begin{tabular}[c]{@{}l@{}}Visual Turn-by-turn Guidance\\ - Gemini Live Single-turn\end{tabular} & -4.86$^{\ddagger}$         & 0.69         & -2.04       & 0.30     & -3.43$^{+}$      & 0.78      & 8.28$^{\ddagger}$        & 1.88       \\
\begin{tabular}[c]{@{}l@{}}Visual Turn-by-turn Guidance\\ - Gemini Live Multi-turn\end{tabular}  & -4.52$^{\dagger}$         & 0.63         & -2.95$^{*}$      & 0.42     & -1.60       & 0.36      & 4.87$^{\ddagger}$        & 1.09       \\
\begin{tabular}[c]{@{}l@{}}Visual Turn-by-turn Guidance \\ - OSPAN\end{tabular}                  & -7.83$^{\ddagger}$         & 1.04         & -4.33$^{\ddagger}$      & 0.60     & 1.98        & 0.57      & 1.91         & 0.55       \\
\begin{tabular}[c]{@{}l@{}}Hands-Free Phone Call \\ - Gemini Live Single-turn\end{tabular}       & -0.47          & 0.06         & 0.18        & 0.02     & -3.88$^{+}$      & 0.6       & 8.53$^{\ddagger}$        & 1.32       \\
\begin{tabular}[c]{@{}l@{}}Hands-Free Phone Call\\ - Gemini Live Multi-turn\end{tabular}         & 0.03           & 0.00         & -0.77       & 0.10     & -1.20       & 0.18      & 3.53$^{+}$        & 0.53       \\
\begin{tabular}[c]{@{}l@{}}Hands-Free Phone Call \\ - OSPAN\end{tabular}                         & -3.54$^{+}$         & 0.40         & -2.33       & 0.28     & 3.18$^{+}$       & 0.75      & -0.04        & 0.01       \\
\begin{tabular}[c]{@{}l@{}}Gemini Live Single-turn\\ - Gemini Live Multi-turn\end{tabular}       & 0.44           & 0.06         & -0.81       & 0.12     & 3.10$^{*}$       & 0.42      & -5.80$^{\ddagger}$       & 0.79       \\
\begin{tabular}[c]{@{}l@{}}Gemini Live Single-turn\\ - OSPAN\end{tabular}                        & -2.42          & 0.34         & -2.04       & 0.30     & 5.95$^{\ddagger}$       & 1.35      & -5.85$^{\ddagger}$       & 1.33       \\
\begin{tabular}[c]{@{}l@{}}Gemini Live Multi-turn\\ - OSPAN\end{tabular}                         & -2.94$^{*}$         & 0.41         & -1.27       & 0.18     & 4.14$^{\dagger}$       & 0.93      & -2.41        & 0.54       \\ \hline
\multicolumn{9}{l}{Note. * p \textless .05, + p \textless .01, † p \textless .001, ‡ p \textless .0001}                                                                                                        
\end{tabular}
\end{table}

\subsection{DRT Response Time}
Results of the repeated measures ANOVA on the DRT response time linear mixed model showed a significant effect for Task, $F$(4, 594.77) = 4.95, $p < $ .001, partial $\eta^2$ = 0.03. The marginal $R^2$ for the response time model was 0.018 and conditional $R^2$ was 0.466. Significant differences in response time were also observed between task conditions, as seen in Figure \ref{fig:DRTdata}. Visual Turn-by-turn Guidance yielded significantly faster response times ($M$ = 0.886, $SE$ = 0.054) than both Gemini Live Multi-turn ($M$ = 0.989, $SE$ = 0.055), $t$(594.08) = -2.95, $p$ = .0299, $d$ = 0.42, and the OSPAN task ($M$ = 1.034, $SE$ = 0.054), $t$(594.40) = -4.33, $p$ = .0002, $d$ = 0.60. Because participants may trade off accuracy for lower response times, interpretations of response times must be considered alongside miss rates. It is particularly important in the current study, where miss rates exceeded 20\% across all task conditions (i.e., hit rates below 80\%), rendering examinations of response times alone insufficient for comparing cognitive load across tasks (ISO, \citeyear{ISO17488}). While response time findings seem to indicate that both the Gemini Live Multi-turn conversational interaction and the OSPAN task were associated with increased cognitive processing demands, as reflected in slower response times, relative to the visual turn-by-turn guidance baseline, miss rate results provide additional nuance. Specifically, the OSPAN benchmark imposed higher cognitive demand than both Gemini Live interaction modes. Taken together, the findings consistently indicate that Gemini Live interaction modes produce cognitive demand comparable to that of hands-free phone call, which is higher than visual turn-by-turn guidance but lower than the OSPAN benchmark. 

\subsection{Gemini Live Multi-turn Analysis}
\begin{figure}[h!]
\begin{center}
\includegraphics[width=1\columnwidth]{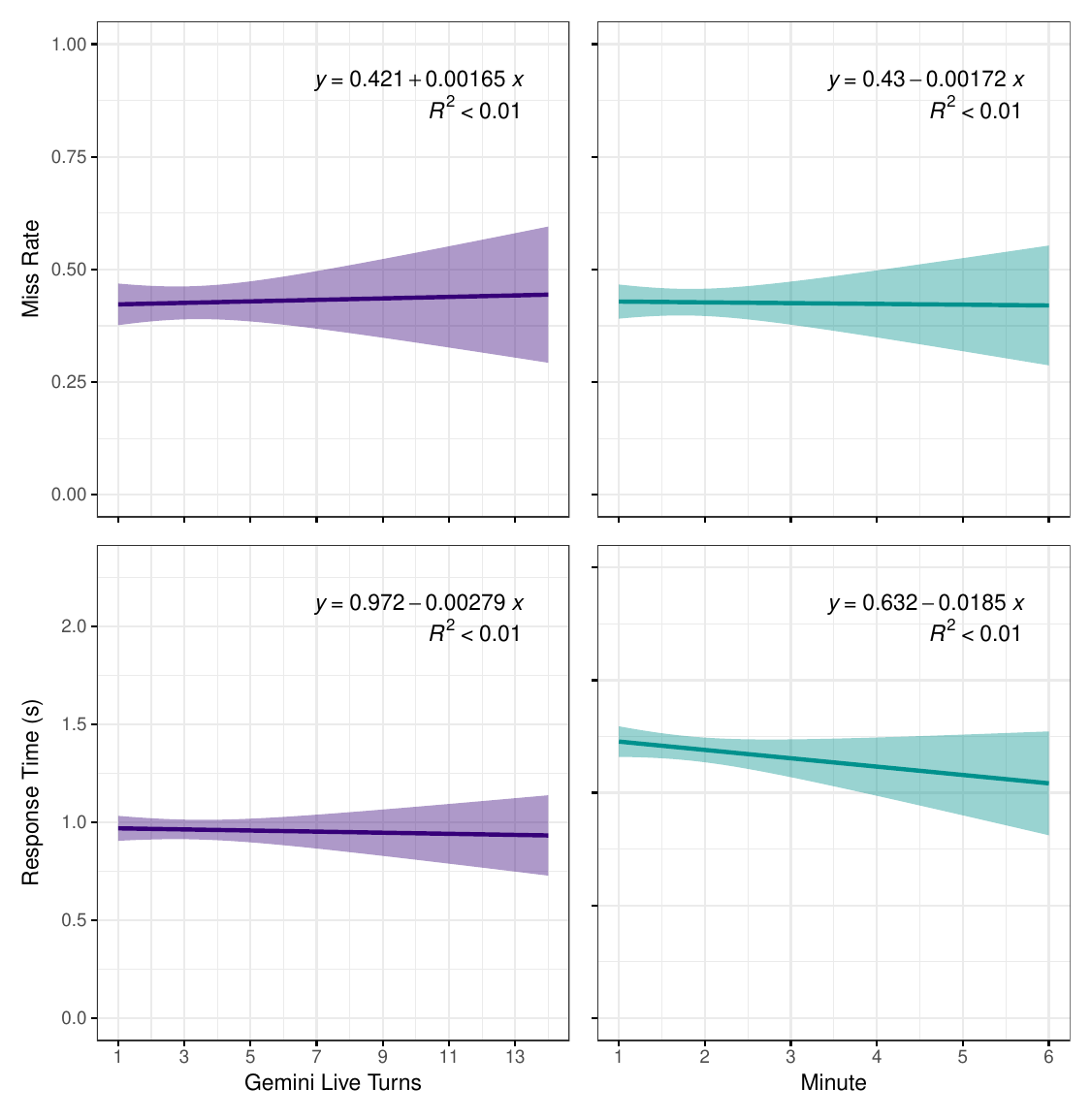}
\end{center}
\caption{Gemini Live Multi-turn duration analysis results.}
\label{fig:GLduration}
\end{figure}

To assess whether cognitive load associated with Gemini Live Multi-turn conversational interactions changed as the interaction progressed, miss rates and response times were analyzed as a function of both the number of conversational turns and elapsed time (by minute). Due to few data points for conversations lasting more than 14 turns or 6 minutes, these comparisons were limited to those time points. As seen in Figure \ref{fig:GLduration}, the data revealed no relationship between miss rate or response time and the number of conversational turns or minutes elapsed. These findings indicate that cognitive load, as indexed by DRT performance, remained stable throughout the course of the Gemini Live Multi-turn interaction, suggesting that extended conversational engagement with the system did not lead to cumulative increases in cognitive demand.

\subsection{Summary of DRT Results}
Analysis of DRT miss rates and response times revealed clear differences in cognitive load across task conditions. Visual Turn-by-turn Guidance consistently imposed the lowest cognitive demand, as evidenced by both the lowest miss rates and the fastest response times. In contrast, all voice-based tasks, including Hands-Free Phone Call and both Gemini Live interaction modes, resulted in significantly higher miss rates than Visual Turn-by-turn Guidance, indicating increased cognitive load. Among all conditions, the OSPAN task produced the highest miss rates and slowest response times, confirming its role as a high-load cognitive benchmark. Notably, Gemini Live Multi-turn interaction and Hands-Free Phone Call elicited similar levels of cognitive demand: higher than Visual Turn-by-turn Guidance, but lower than OSPAN. Importantly, exploratory analyses of cognitive load across the duration of Gemini Live interactions indicated that miss rates and response times remained relatively stable over time, suggesting that drivers did not experience cumulative increases in cognitive workload as conversations progressed. These findings suggest that while conversational AI and phone-based voice interactions do increase cognitive workload relative to an on-going low visual-demand task, their demands remain below those of complex working memory challenges.

\subsection{Mean Glance Duration (MGD)}
\begin{figure}[h!]
\begin{center}
\includegraphics[width=1\columnwidth]{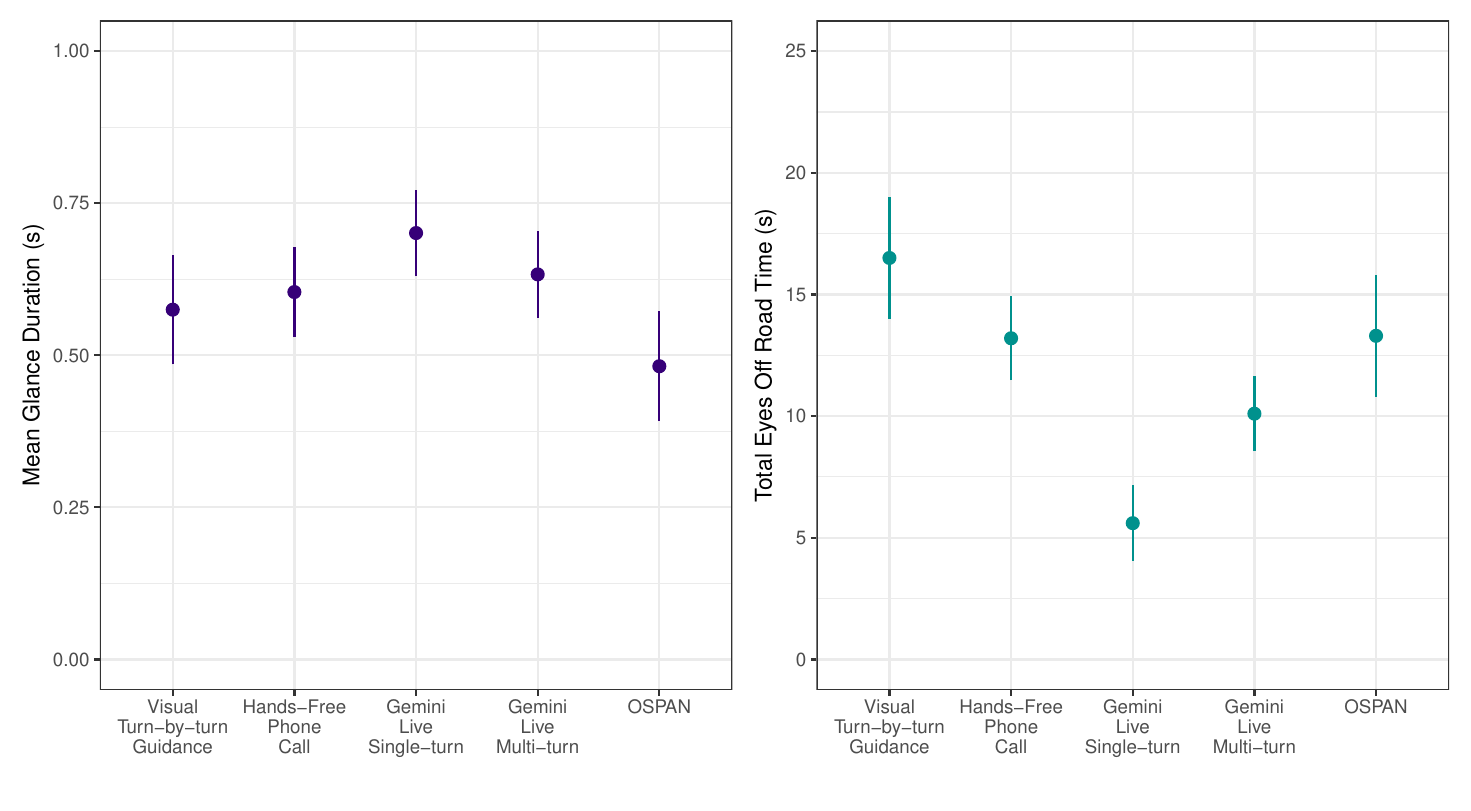}
\end{center}
\caption{Eye glance estimated marginal means and 95\% Confidence Intervals.}
\label{fig:glancedata}
\end{figure}

Results of the repeated measures ANOVA on the MGD linear mixed model showed a significant effect for Task, $F$(4, 312.32) = 11.02, $p < $ .0001, partial $\eta^2$ = 0.12. The marginal $R^2$ for the MGD model was 0.065 and conditional $R^2$ was 0.504. As seen in Figure \ref{fig:glancedata}, all task conditions were well below the 2-second MGD criterion established by NHTSA and the Alliance (NHTSA, \citeyear{nhtsa2013}, \citeyear{nhtsa2014}; Alliance, \citeyear{alliance2006}). Even so, there were several significant differences among the task conditions. Participants exhibited significantly longer MGDs in the Gemini Live Single-turn condition ($M$ = 0.701, $SE$ = 0.035) compared to both Visual Turn-by-turn Guidance ($M$ = 0.575, $SE$ = 0.045), $t$(312.19) = -3.43, $p$ = .0047, $d$ = 0.78, and Hands-Free Phone Call ($M$ = 0.604, $SE$ = 0.037), $t$(312.57) = -3.88, $p$ = .001, $d$ = 0.60. Additionally, the Gemini Live Single-turn condition yielded significantly longer MGDs scores than the Gemini Live Multi-turn condition ($M$ = 0.633, $SE$ = 0.035), $t$(312.64) = 3.10, $p$ = .011, $d$ = 0.42. The OSPAN task ($M$ = 0.482, $SE$ = 0.045) produced significantly shorter MGDs than Hands-Free Phone Call, $t$(311.86) = 3.18, $p$ = .0098, $d$ = 0.75, Gemini Live Single-turn, $t$(312.19) = 5.95, $p <$ .0001, $d$ = 1.35, and Gemini Live Multi-turn, $t$(312.21) = 4.14, $p$ = .0004, $d$ = 0.93. These results indicate that Gemini Live Single-turn generally elicited longer MGDs than other conditions, while OSPAN was associated with the shortest MGDs among the tasks.

\begin{table}[h!]
\caption{Estimated marginal means table for off-road and on-road MGD.}
\label{table4}
\begin{tabular}{lcccccccccc}
\hline
                                                                        & \multicolumn{5}{c}{Off-Road MGD}  & \multicolumn{5}{c}{On-Road MGD}       \\ \hline
Task                                                                    & Est. & SE   & df    & LCI  & UCI  & Est.  & SE   & df     & LCI   & UCI   \\ \hline
\begin{tabular}[c]{@{}l@{}}Visual Turn-by-\\ turn Guidance\end{tabular} & 0.58 & 0.05 & 82.92 & 0.48 & 0.67 & 10.83 & 1.19 & 497.55 & 7.48  & 14.18 \\
\begin{tabular}[c]{@{}l@{}}Hands-Free \\ Phone Call\end{tabular}        & 0.60 & 0.04 & 37.07 & 0.53 & 0.68 & 9.78  & 0.76 & 173.25 & 7.63  & 11.94 \\
\begin{tabular}[c]{@{}l@{}}Gemini Live \\ Single-turn\end{tabular}      & 0.70 & 0.04 & 31.30 & 0.63 & 0.77 & 3.92  & 0.68 & 118.28 & 1.97  & 5.87  \\
\begin{tabular}[c]{@{}l@{}}Gemini Live \\ Multi-turn\end{tabular}       & 0.63 & 0.04 & 29.88 & 0.56 & 0.70 & 7.41  & 0.64 & 95.37  & 5.57  & 9.25  \\
OSPAN                                                                   & 0.48 & 0.05 & 82.92 & 0.39 & 0.57 & 19.14 & 1.19 & 497.55 & 15.79 & 22.49 \\ \hline
\end{tabular}
\end{table}

To further explore glance behavior during voice-based tasks, the estimated marginal means for on-road and off-road glances within each task condition were examined. As shown in Table \ref{table4}, the mean durations of on-road glances were consistently much longer than those of off-road glances for each of the five tasks. The differences were significant for each of the five tasks: Visual Turn-by-turn Guidance $t$(630.07) = 6.44, $p <$ .0001, $d$ = 1.86, Hands-free Phone Call $t$(630.07) = 9.94, $p <$ .0001, $d$ = 1.67, Gemini Live Single-turn $t$(630.07) = 4.06, $p$ = .002, $d$ = 0.58, Gemini Live Multi-turn $t$(630.07) = 9.32, $p <$ .0001, $d$ = 1.22, and OSPAN $t$(630.07) = 11.72, $p <$ .0001, $d$ = 3.38. The clear separation between off-road and on-road MGDs highlights how drivers interleave longer on-road glances between off-road glances to the secondary task device for both voice-based tasks, and in this case, even a low-demand visual task.

\subsection{Total Eyes Off Road Time (TEORT)}
Results of the repeated measures ANOVA on the TEORT linear mixed model showed a significant effect for Task, $F$(4, 313.02) = 29.67, $p <$ .0001, partial $\eta^2$ = 0.27. The marginal $R^2$ for the TEORT model was 0.229 and conditional $R^2$ was 0.355. Post-hoc paired comparisons revealed significant differences in TEORT across task conditions (see Figure \ref{fig:glancedata}). Visual Turn-by-turn Guidance produced the longest TEORT ($M$ = 16.50, $SE$ = 1.29), which was significantly longer than both Gemini Live Single-turn ($M$ = 5.60, $SE$ = 0.77), $t$(312.58) = 8.28, $p <$ .0001, $d$ = 1.88, and Gemini Live Multi-turn ($M$ = 10.14, $SE$ = 0.75), $t$(312.59) = 4.87, $p <$ .0001, $d$ = 1.09. Similarly, Hands-Free Phone Call ($M$ = 13.21, $SE$ = 0.86) resulted in significantly longer TEORT than Gemini Live Single-turn, $t$(313.85) = 8.53, $p <$ .0001, $d$ = 1.32, and Gemini Live Multi-turn, $t$(314.18) = 3.53, $p$ = .0024, $d$ = 0.53. Direct comparison between the two Gemini Live conditions revealed that the Multi-turn version ($M$ = 10.13, $SE$ = 0.75) was associated with significantly longer TEORTs than the Single-turn version ($M$ = 5.60, $SE$ = 0.77), $t$(314.19) = -5.80, $p <$ .0001, $d$ = 0.79. Notably, the inherently shorter task duration for Gemini Live Single-turn likely contributed to its lower TEORT relative to the other conditions. Finally, Gemini Live Single-turn also yielded significantly shorter TEORT compared to the OSPAN task ($M$ = 13.26, $SE$ = 1.29), $t$(312.58) = -5.85, $p <$ .0001, $d$ = 1.33. Collectively, these results indicate that both Gemini Live interaction modes, particularly the Single-turn condition, were associated with lower TEORT compared to the other task conditions. The longer TEORT observed for the Visual Turn-by-turn Guidance and OSPAN conditions is not surprising given the 5-minute data collection window for the turn-by-turn guidance task and 6-minutes for OSPAN. In contrast, Gemini Live Multi-turn averaged just over 2 minutes, and Single-turn only 35 seconds, which likely contributed to their lower TEORT values.

\subsection{Summary of Glance Data}
Analysis of visual attention demand revealed significant differences across task conditions. All task conditions were well below the 2-second MGD criterion established by NHTSA and the Alliance (NHTSA, \citeyear{nhtsa2013}, \citeyear{nhtsa2014}; Alliance, \citeyear{alliance2006}), indicating no concern for mean glance durations. Interpretation of TEORT results was more nuanced. Both Gemini Live conditions had mean TEORT values below 12 seconds, yet only the Single-turn condition met the NHTSA criterion that at least 85 percent of participants had mean TEORT values below 12 seconds. When applying the Alliance’s 20-second TEORT criterion, Gemini Live Multi-turn and Hands-Free Phone Call, along with Single-turn, had passing TEORT results. The longer TEORTs observed for Visual Turn-by-turn Guidance and OSPAN were not surprising given their longer data-collection windows (5 minutes for turn-by-turn guidance and 6 minutes for OSPAN), whereas Gemini Live Multi-turn averaged just over 2 minutes and Single-turn only 35 seconds, which likely contributed to their lower TEORT values.

Further examination of glance patterns revealed a clear separation between off-road and on-road mean glance durations. On-road MGDs for Visual Turn-by-turn Guidance, Hands-Free Phone Call, Gemini Live Multi-turn, and OSPAN all exceeded 6 seconds, a duration found sufficient for rebuilding situation awareness \parencite{meyer2022impact,seaman2021vehicle}.

Taken together, these findings show that all task conditions maintained mean glance durations well below the established 2-second threshold, while TEORT outcomes were influenced by the longer durations of certain tasks, yet Hands-Free Phone Call and both Gemini Live conditions met the Alliance’s 20-second criterion. The predominance of on-road MGDs exceeding 6 seconds indicates drivers consistently dedicated sufficient time between device glances to rebuild roadway situation awareness, such that the overall visual demand across the five tasks did not rise to a level of concern.  

\subsection{Subjective Ratings of Demand and Performance}
At the end of data collection, participants were asked a series of subjective feedback questions, including how much they agree with statements describing task demands and performance (on a scale of “1”, “strongly disagree” to “5”, “strongly agree”). According to median values provided in Table \ref{table5}, overall ratings of effort, frustration, and various types of demands were low for the Visual Turn-by-turn Guidance, hands-free phone call, and Gemini Live tasks. While participants also strongly disagreed that the OSPAN task was physically demanding, participants tended to agree that the OSPAN task was mentally and temporally demanding, with some rating the task as requiring effort. Relatedly, while participants were generally satisfied with their performance on other tasks, some disagreed with being satisfied with their performance in OSPAN. Regarding how distracting these tasks were, participants overall strongly disagreed that any of the tasks were visually distracting. While participants tended to agree that they remained aware of the road while completing all of the four tasks, some disagreed that they would do the OSPAN task while driving. In contrast, participants generally strongly agreed that they would do Visual Turn-by-turn Guidance, hands-free phone calls, and Gemini Live tasks while driving.

\begin{table}[h!]
\caption{Medians of subjective ratings by metric and task.}
\label{table5}
\begin{tabular}{cclcccc}
\hline
\multicolumn{2}{c}{Metrics}                                                                                                                                            & Scale Item                                                                                                                                                    & \begin{tabular}[c]{@{}c@{}}Visual \\ Turn-by-\\ turn   \\ Guidance\end{tabular} & \begin{tabular}[c]{@{}c@{}}Hands-\\ Free \\ Phone \\ Call\end{tabular} & \begin{tabular}[c]{@{}c@{}}Gemini \\ Live\end{tabular} & OSPAN                \\ \hline
\multicolumn{2}{c}{Effort}                                                                                                                                             & \begin{tabular}[c]{@{}l@{}}I thought this task \\ was difficult to \\ accomplish.\end{tabular}                                                                & 1.0                                                                             & 1.0                                                                    & 1.0                                                    & 3.0                  \\ \hline
\multicolumn{2}{c}{Frustration}                                                                                                                                        & \begin{tabular}[c]{@{}l@{}}I felt irritated and/\\ or stressed during \\ this task.\end{tabular}                                                              & 1.0                                                                             & 1.0                                                                    & 1.0                                                    & 2.0                  \\ \hline
\multicolumn{2}{c}{Mental Demand}                                                                                                                                      & \begin{tabular}[c]{@{}l@{}}I felt this task \\ required a lot of \\ mental effort (i.e., \\ thinking, decision \\ making, \\ remembering, etc.).\end{tabular} & 1.0                                                                             & 1.0                                                                    & 1.0                                                    & 4.5                  \\ \hline
\multicolumn{2}{c}{Physical Demand}                                                                                                                                    & \begin{tabular}[c]{@{}l@{}}I thought this task \\ was physically \\ strenuous (i.e., \\ hard to reach the \\ controls).\end{tabular}                          & 1.0                                                                             & 1.0                                                                    & 1.0                                                    & 1.0                  \\ \hline
\multicolumn{2}{c}{Temporal Demand}                                                                                                                                    & \begin{tabular}[c]{@{}l@{}}I felt rushed while \\ trying to complete \\ this task.\end{tabular}                                                               & 1.0                                                                             & 1.0                                                                    & 1.0                                                    & 4.0                  \\ \hline
\multicolumn{2}{c}{\begin{tabular}[c]{@{}c@{}}Performance \\ Satisfaction\end{tabular}}                                                                                & \begin{tabular}[c]{@{}l@{}}I am satisfied with \\ my performance of \\ this task.\end{tabular}                                                                & 5.0                                                                             & 5.0                                                                    & 4.5                                                    & 3.0                  \\ \hline
\multirow{5}{*}{\begin{tabular}[c]{@{}c@{}}Perceived \\ Distraction\end{tabular}} & \multirow{2}{*}{Visual}                                                            & \multirow{2}{*}{\begin{tabular}[c]{@{}l@{}}The feature was \\ visually distracting.\end{tabular}}                                                             & \multirow{2}{*}{1.0}                                                            & \multirow{2}{*}{1.0}                                                   & \multirow{2}{*}{1.5}                                   & \multirow{2}{*}{1.0} \\
                                                                                  &                                                                                    &                                                                                                                                                               &                                                                                 &                                                                        &                                                        &                      \\ \cline{2-7} 
                                                                                  & \begin{tabular}[c]{@{}c@{}}Road \\ Awareness\end{tabular}                          & \begin{tabular}[c]{@{}l@{}}I remained aware \\ of the road when \\ doing this task.\end{tabular}                                                              & 5.0                                                                             & 4.0                                                                    & 4.0                                                    & 4.0                  \\ \cline{2-7} 
                                                                                  & \multirow{2}{*}{\begin{tabular}[c]{@{}c@{}}Would do \\ while driving\end{tabular}} & \multirow{2}{*}{\begin{tabular}[c]{@{}l@{}}I would do this \\ task while driving.\end{tabular}}                                                               & \multirow{2}{*}{5.0}                                                            & \multirow{2}{*}{5.0}                                                   & \multirow{2}{*}{5.0}                                   & \multirow{2}{*}{2.5} \\
                                                                                  &                                                                                    &                                                                                                                                                               &                                                                                 &                                                                        &                                                        &                      \\ \hline
\end{tabular}
\end{table}

\begin{figure}[h!]
\begin{center}
\includegraphics[width=1\columnwidth]{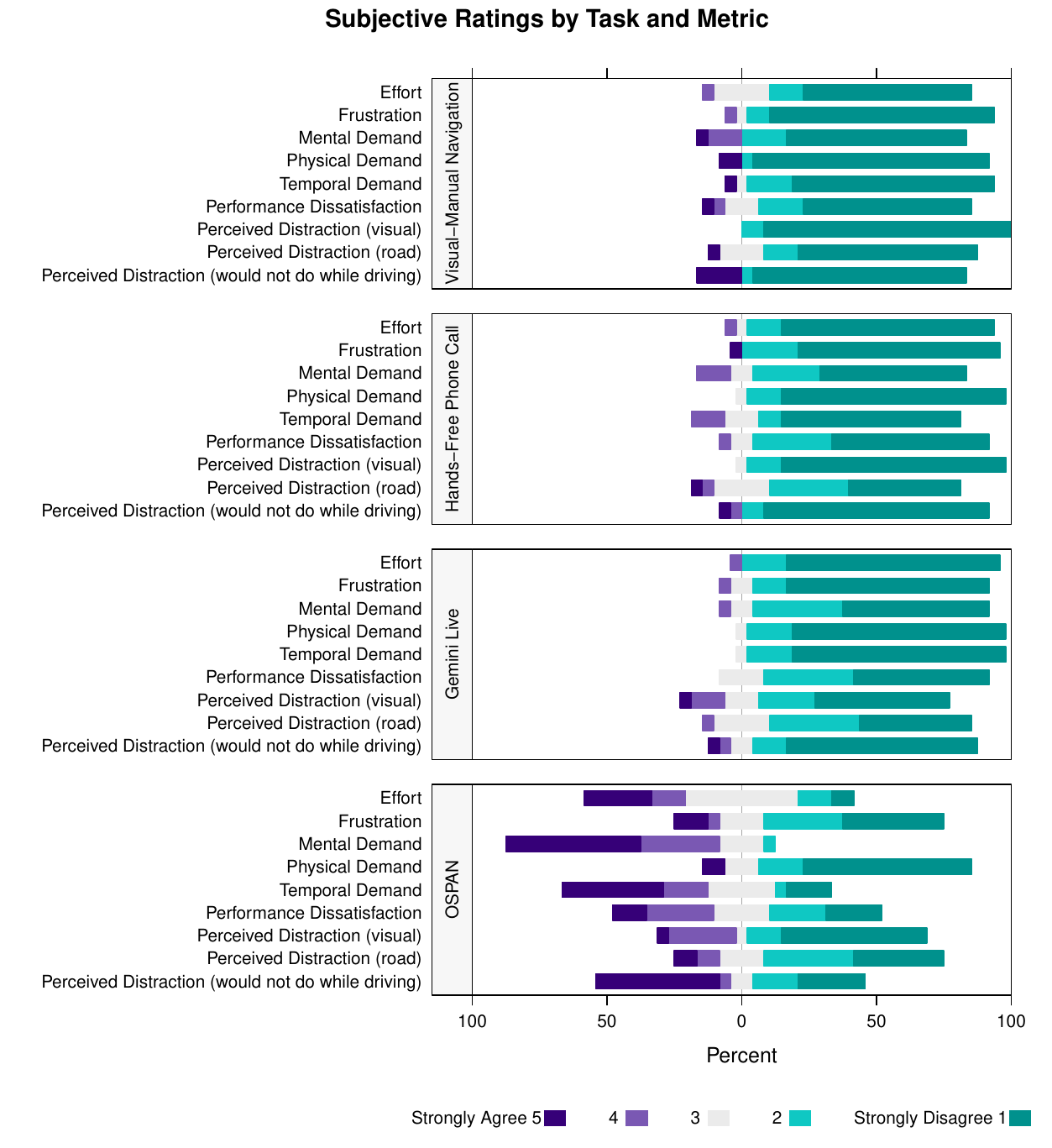}
\end{center}
\caption{Subjective Ratings results by Task and Metric.}
\label{fig:subjdata}
\end{figure}

Analysis of the ratings, in terms of the percentage of participants who provided ratings at each point of the one to five scale, provided similar results (see Figure \ref{fig:subjdata}). Metric items that originally had positive valence were flipped along the scale for this analysis, such that higher ratings always indicated lower demands and higher satisfaction in performance of tasks.

\subsection{Subjective Ratings of Gemini Live}
Subjective ratings specifically about experiences with using Gemini Live revealed that participants generally strongly agreed with statements of satisfaction with the experience, the technology’s ease of use, usefulness, clarity, and reliability. Participants rated their experience by how much they agreed with various statements describing their interactions with Gemini Live on a scale of “1” for “strongly disagree” to “5” for “strongly agree.” Participants finally also rated, using a five-point scale, how likely they were to use Gemini Live on Android Auto Projected again (from “1” for “very unlikely,” to “5” for “very likely”). See Table \ref{table6} for median ratings by each metric.

\begin{table}[h!]
\caption{Median ratings of various aspects of participant experience interacting with Gemini Live.}
\label{table6}
\begin{tabular}{lll}
\hline
Metric                                                                               & Scale Item                                                                                                                                                                                                                                            & Median \\ \hline
Satisfaction                                                                         & \begin{tabular}[c]{@{}l@{}}I am satisfied with my experience   using Gemini Live on \\ Android Auto Projected.\end{tabular}                                                                                                                           & 5.0    \\ \hline
Ease of Use                                                                          & Using Gemini Live on Android Auto Projected was easy.                                                                                                                                                                                                 & 4.5    \\ \hline
Usefulness                                                                           & \begin{tabular}[c]{@{}l@{}}Gemini Live’s responses on Android   Auto Projected \\ were useful.\end{tabular}                                                                                                                                           & 5.0    \\ \hline
Clarity                                                                              & I found Gemini Live’s responses to   be easy to follow.                                                                                                                                                                                               & 5.0    \\ \hline
Reliability                                                                          & \begin{tabular}[c]{@{}l@{}}Gemini Live on Android Auto   Projected successfully \\ accomplished what I asked it to do.\end{tabular}                                                                                                                   & 5.0    \\ \hline
\begin{tabular}[c]{@{}l@{}}Mental Effort \\ (Amount)\end{tabular}                    & \begin{tabular}[c]{@{}l@{}}Interacting with Gemini Live while   driving required \\ a lot of mental effort (attention, thinking, decision \\ making, remembering, etc.).\end{tabular}                                                                 & 2.0    \\ \hline
\begin{tabular}[c]{@{}l@{}}Mental Effort \\ (Compared to \\ Phone Call)\end{tabular} & \begin{tabular}[c]{@{}l@{}}Interacting with Gemini Live while   driving required \\ the same mental effort (attention, thinking, decision \\ making, remembering, etc.) as holding a hands-free \\ cellphone conversation while driving.\end{tabular} & 3.0$^{*}$   \\ \hline
Would use again                                                                      & \begin{tabular}[c]{@{}l@{}}How likely are you to use Gemini   Live on Android\\ Auto Projected again?\end{tabular}                                                                                                                                    & 5.0    \\ \hline
\multicolumn{3}{l}{\textit{\begin{tabular}[c]{@{}l@{}}* Of all participants who disagreed that interacting with Gemini Live while driving \\ required the same mental effort as a phone conversation (rating \textless 3; n = 11), 6 \\ described Gemini Live as requiring more effort.\end{tabular}}}                                               
\end{tabular}
\end{table}

\subsection{Summary of Subjective Ratings Data}
Participants’ subjective ratings generally aligned with the objective results, indicating low perceived effort, frustration, and demand for visual turn-by-turn guidance, hands-free phone calls, and both Gemini Live conditions. In contrast, the OSPAN task was consistently rated as mentally and temporally demanding, with some participants reporting higher effort and lower satisfaction with their performance. Across all tasks, participants strongly disagreed that the activities were visually distracting and generally agreed they remained aware of the road. Most participants indicated they would perform visual turn-by-turn guidance, hands-free phone calls, and Gemini Live tasks while driving, whereas many expressed reluctance to attempt the OSPAN task in real driving scenarios. Ratings specific to Gemini Live were highly positive, with strong agreement on ease of use, clarity, usefulness, and reliability, and participants reported high likelihood of using the technology again in future driving contexts.

\section{Discussion}\label{sec:discussion}

This study provides a comprehensive evaluation of cognitive load and visual attention demand associated with conversational AI (Gemini Live Single-turn and Multi-turn), hands-free phone call, visual turn-by-turn guidance, and a high-load OSPAN task during on-road driving. At a high level, the results show that Gemini Live interactions and hands-free phone calls imposed similar levels of cognitive load, higher than visual turn-by-turn guidance but lower than OSPAN, while all voice-based tasks maintained low visual attention demand, with mean glance durations well below NHTSA’s 2-second criterion. Both Gemini Live conditions had mean TEORT values below 12 seconds, but only the Single-turn condition met the NHTSA criterion for 85 percent of participants. When using the Alliance’s 20-second TEORT criterion, Gemini Live Multi-turn and Hands-Free Phone Call also passed. However, the on-road MGDs for all tasks except Gemini Live Single-turn exceeded 6 seconds, indicating drivers had sufficient time between device glances to rebuild roadway situation awareness. Subjective ratings of workload and distraction generally aligned with objective measures, and exploratory analyses revealed that cognitive load during Gemini Live Multi-turn interactions remained stable over time.

The first hypothesis predicted that Gemini Live interactions and hands-free phone calls would impose similar levels of cognitive load, greater than visual turn-by-turn guidance but less than the high-load OSPAN task. Consistent with this prediction, both Gemini Live conditions and hands-free phone calls produced cognitive load levels between those observed for visual turn-by-turn guidance and the OSPAN task. This was reflected in both DRT miss rates and response times, with Gemini Live Multi-turn and hands-free phone call yielding significantly higher values than visual turn-by-turn guidance, but significantly lower than OSPAN. These findings are in line with prior research showing that hands-free phone calls and voice-based digital assistants increase cognitive demand relative to simple navigation tasks, but do not approach the high workload imposed by complex working memory challenge \parencite{monk2023visual,strayer2013measuring,bond2025chatgpt}.

The second hypothesis proposed that all task conditions would result in MGDs well below NHTSA’s and the Alliance’s 2-second criterion, indicating minimal visual attention demand. As anticipated, all conditions, including both Gemini Live modes, produced mean glance durations well below this threshold, suggesting no practical concern for prolonged glances away from the roadway. While there were statistical differences in MGD across conditions, these differences were small in magnitude and not meaningful from a safety perspective, echoing findings from \textcite{monk2023visual} and \textcite{mehler2016multi}.

The third hypothesis anticipated that subjective ratings of workload and distraction would align closely with objective measures, reflecting consistent patterns across task types. Subjective ratings of workload and distraction closely tracked the objective measures. Participants reported low effort, frustration, and distraction for Gemini Live, hands-free phone calls, and visual turn-by-turn guidance, while the OSPAN task was consistently rated as more mentally and temporally demanding, with lower satisfaction. These results reinforce the value of combining subjective and objective metrics for a holistic assessment of driver workload \parencite{monk2023visual,von2022subjective,biondi2024adopting}.

The exploratory question examined whether cognitive load during extended Gemini Live Multi-turn interactions remained stable or changed over time. Exploratory analyses revealed that cognitive load, as indexed by DRT miss rates and response times, remained stable throughout the course of Gemini Live Multi-turn interactions. There was no evidence of cumulative increases in cognitive workload as conversations progressed, whether measured by number of conversational turns or elapsed time. In addition, the comparison between off-road and on-road MGDs revealed that drivers dedicated longer glances to the road ahead between brief glances to the device, especially during voice-based tasks, which suggests these conversational interfaces pose less visual attention demand. Taken together, this suggests that, within the timeframes studied, extended engagement with conversational AI does not lead to escalating cognitive demand. 

A notable pattern in the present findings is that the DRT was most sensitive to differences at the extremes of the cognitive load spectrum, clearly distinguishing between low-load (visual turn-by-turn guidance) and high-load (OSPAN) conditions, while showing less diagnostic sensitivity among tasks such as Gemini Live and hands-free phone call that fall between. This pattern is consistent with recent work by \textcite{biondi2024adopting}, who found that the DRT robustly detects large differences in cognitive workload (e.g., between different road types or vehicles) but is less effective at discriminating between tasks that impose moderate or similar levels of demand. As \textcite{biondi2024adopting} and others have noted, this limitation should be considered when interpreting DRT results in studies comparing modern in-vehicle technologies, where differences may be subtle and multifactorial.

The present findings have important implications for the introduction of conversational AI into vehicles. The results suggest that, when implemented thoughtfully, conversational agents like Gemini Live do not impose excessive cognitive or visual demands on drivers, especially when compared to traditional benchmarks such as hands-free phone call or high-load cognitive tasks. Across all tasks, including both Gemini Live conditions, mean glance durations remained well below the established 2-second threshold, indicating minimal visual attention demand. While only Gemini Live Single-turn met the NHTSA TEORT criterion, Gemini Live Multi-turn and hands-free phone call also satisfied the Alliance’s TEORT threshold. Notably, for tasks that did not pass the NHTSA TEORT recommendation, average on-road glance durations exceeded the level shown to rebuild situation awareness, suggesting that overall visual attention demand did not rise to a level of concern \parencite{meyer2022impact,seaman2021vehicle}. In addition, cognitive load remained moderate and stable over time. These findings align with recent simulator and on-road studies showing that voice-based interfaces and advanced conversational agents are generally well-accepted by users and can help minimize distraction potential by providing an auditory-only substitution for in-vehicle visual-manual tasks \parencite{bond2025chatgpt,monk2023visual}. Research indicates that replacing these visual-manual tasks with voice-based interaction significantly reduces eyes-off-road time and manual interference, improving lane-keeping and overall driving performance \parencite{chiang2005comparison,shutko2009,monk2023visual}. 

Interpreted through the lens of CASA and machine heuristic frameworks, the comparable cognitive demands of Gemini Live and hands-free phone call suggest that drivers may respond to conversational AI in ways similar to human interlocutors \parencite{reeves1996media,lee2024minding,henry-abion2025}. In-vehicle conversational agents may provide benefits to drivers such as increased alertness offering a substitution for passenger conversation that slightly increases cognitive load as a means to counteract fatigue or underload in low demand driving environments \parencite{gershon2009effects}. This has important implications for the design of future in-vehicle assistants, indicating that anthropomorphic qualities and adaptive dialogue may foster engagement and trust without imposing excessive cognitive or visual demands. As conversational agents become more sophisticated, capable of sustaining multi-turn dialogue, expressing empathy, and adapting to user preferences, it is important to recognize how these human-like features can influence user expectations and interaction patterns.

However, the study also highlights the importance of continued vigilance. While no evidence was found for escalating cognitive load during extended multi-turn interactions, and subjective ratings were favorable, the DRT’s limited sensitivity to moderate differences means that subtle or context-dependent effects cannot be ruled out. As conversational AI systems continue to improve in performance and quality and are integrated with a broader range of vehicle functions, ongoing evaluation using multiple complementary measures will be critical to monitor their evolving impact on cognitive and visual demands across diverse driving contexts and user populations.

\subsection{Limitations}\label{sec:limitations}

Several limitations should be considered when interpreting the findings of this study. First, participants exhibited variable willingness and ability to sustain conversations with Gemini Live. Although all participants were screened for prior experience with voice-based conversational agents, some had only interacted with systems structured similarly to the Single-turn mode and initially struggled with the sustained dialogue required in Multi-Turn tasks. This occasionally resulted in participants speaking over Gemini Live, causing the system to begin a new response before finishing the prior dialogue. Most participants overcame this challenge during practice rounds and adapted to the flow of Multi-Turn conversations during the driving trials.

Second, the choice of a driving route that minimized traffic and stoplights inadvertently introduced connectivity issues. In areas with suboptimal cell phone service, interactions with Gemini Live were sometimes interrupted, requiring the experimenter to reset the system before participants could continue. These interruptions may have affected the continuity and naturalness of the conversational AI experience. In addition, the test route may not be representative of other non-highway driving conditions.

Third, as is typical in on-road studies, some participants encountered unexpected roadway events that required their full attention. For example, a student driver abruptly stopped in front of the experimental vehicle, prompting the experimenter to pause the Gemini Live and Hands-Free Phone Call tasks until the participant felt comfortable resuming. Such real-world events may have influenced task engagement and data collection.

Fourth, participants were instructed to complete tasks in a manner that prevented them from choosing when to interact with the system. In real-world use, drivers choose when to initiate interactions, such as when stopped at a red light or during low demand driving segments. The study route was designed to minimize traffic, but increased congestion or different driving environments could alter how and when users engage with Gemini Live.

Fifth, variability in drive completion times and connectivity issues led to reduced time available for some participants to complete all tasks. In these cases, only practice trial data were available for analysis, potentially limiting the completeness and consistency of the dataset.

Finally, individual differences that were not fully accounted for, such as participant talkativeness and familiarity with the driving route, may have influenced task performance and engagement. Future studies should consider these factors and explore a broader range of driving contexts, user populations, and conversational AI features to enhance generalizability.

\subsection{Summary}\label{sec:future}

This study provides a comprehensive assessment of cognitive and visual attention demands across conversational AI, hands-free phone use, visual turn-by-turn guidance, and a high-load working memory task during on-road driving. Overall, Gemini Live interactions and hands-free phone calls imposed similar levels of cognitive load, while all voice-based tasks maintained low visual attention demand, with glance durations well below established thresholds. Subjective ratings generally mirrored objective measures, and exploratory analyses indicated that cognitive load during extended Gemini Live Multi-turn interactions remained stable over time. Notably, the clear separation between off-road and on-road mean glance durations highlights how drivers tend to dedicate longer glances to the road ahead between off-road glances to the secondary task device, even for voice-based tasks and a low-demand visual task, helping maintain roadway situation awareness and rendering the eyes-off-road time findings less consequential. These findings underscore the importance of continued research as conversational AI systems continue to improve in quality, performance, and support more naturalistic, seamless interactions for drivers as they are increasingly integrated with vehicle functions. Future work can incorporate diverse, complementary metrics to monitor how evolving designs influence driver workload and attention across varied contexts and user populations.

\section{Funding}
This work was supported by Google, Inc., Mountain View, CA. The funder played a role in the conceptualization of the study, provided access to Gemini Live integrated with Android Auto, and contributed the decision to submit the manuscript for publication.

% \section{CRediT authorship contribution statement}
% Chris Monk: Conceptualization, Methodology, Investigation, Writing - original draft, Formal analysis, Funding acquisition, Project administration. Allegra Ayala: Methodology, Data collection, Writing – original draft. Christine Yu: Methodology, Data Reduction, Data Analysis, Writing – review \& editing. Greg Fitch: Resources, Conceptualization. Dara Gruber: Resources, Conceptualization.

% \section{Declaration of Competing Interest}
% Chris Monk, Allegra Ayala, and Christine Yu have no known competing interests or personal relationships that could have appeared to influence the work reported in this paper. Greg Fitch and Dara Gruber are employees and own stock in Google, Inc.

{\singlespacing
\printbibliography
}

\end{document}